\documentclass[11pt,twoside]{article}

\usepackage{asp2006_hotcool}
\usepackage{epsf}
\usepackage{psfig}
\usepackage{lscape}
\usepackage{graphicx}
\newcommand{\Ms}{M$\sun$}
\newcommand{\til}{~}
\begin{document}
\title{Dust formation in massive stars and their explosive ends}   
\author{I. Cherchneff}   
\affil{Departement Physik, Universit{\"a}t Basel, Klingelbergstrasse 82, CH-4056 Basel, Switzerland.}    

\begin{abstract} 
We review the observational evidence for dust formation in Wolf-Rayet binary systems and in Type II Supernova ejecta. Existing theoretical models describing the condensation of solids in carbon-rich Wolf-Rayet stars and in Supernovae close by and at high redshift are discussed. We describe new modeling of carbon- and oxygen-based grain nucleation using a chemical kinetic approach applied to the ejecta of massive pair-instability Supernovae in the early universe. Finally, dust formation processes in colliding wind regions of WC binary systems are discussed.
\end{abstract}


\section{Introduction}
 Late stages of evolution of massive stars are interesting space laboratories to study the formation of molecules and dust under inhospitable conditions. Their study has been rekindled by the potential important role played by very massive stars in the first enrichment of gas in the early universe. Stars with mass larger than 8 \Ms~ evolve to the supergiant phase on short time scales ($\sim$ 10$^6$ years) compared to their low-mass counterparts. Dependent of the initial mass of the star on the ZAMS, stars in the lower part of the mass distribution will evolve as Supergiants whereas very massive objects with ZAMS mass of $\sim$ 40 \Ms\til will become Wolf-Rayet (WR) stars. Both Supergiants and Wolf-Rayet stars end up their stellar life explosively, in the form of Type II and Type Ib/c Supernovae (hereafter SNe), respectively. Dust is observed in the most advanced evolutionary stage of WR stars, the carbon-rich WRs (hereafter WCs), where strong mass loss has stripped out the entire hydrogen envelope and a helium/carbon-rich photosphere is exposed. A carbon-based solid is expected to form in WC winds owing to the photospheric chemical composition. Evidence is growing that those rare massive stars are part of a binary system with a OB star companion.  
Type IIp supernovae ('p' stands for the plateau phase of their light curve) are by far the most common SNe. The debate is still wide open as to whether SNe are significant dust makers in our local universe, despite the evidence of dust formation in situ in several SN ejecta, amongst which SN1987A in the Large Magellanic Cloud. This issue becomes of paramount importance for explaining the presence of large amount of dust in QSOs in the early universe. At z $>$ 6, only very massive stars (M$_{\star} > $100 \Ms) are expected to form because of evolutionary time constraints. It is thus conjectured that their explosive end as pair-instability SNe (hereafter PISNe) is a potential locus for dust formation when the universe was less than 1 Gyr old. 

All those massive circumstellar media, although different in origin, show the following common parameters: a hydrogen-poor/helium-rich gas composition, a wind/ejecta chemical composition rich in heavy elements and metals  (i.e., O, C, Si, S, Mg, Al, Fe and Ca), large flow velocities ($\sim$ 2000 km s$^{-1}$) and very high gas temperatures to start with (20000 - 45000~K). How molecules and dust can form and withstand the extreme physical conditions met in those environments is still an open question. In the following, we shall concentrate on two different types of late, massive, dusty circumstellar media, WC winds and PISNe ejecta in the early universe. We first discuss observational evidences for dust and review the various types of possible condensates in those environments as well as existing studies on the subject. We finally briefly present new results on dust nucleation in PISN ejecta with possible implications to WC colliding winds. 
 
  \section{Evidence for dust in Wolf-Rayet stars}
 First evidence for IR excess due to the existence of dust in the winds of late WCs is given by Allen et al. (1972). Several subsequent observational studies pinpoint that dusty WC stars are of two kinds, permanent and periodic dust makers, and are part of binary systems where the companion is a hydrogen-rich OB star (for recent reviews, see Crowther 2003, Marchenko \& Moffat 2007, Williams 2008). The region where the dense hot wind of the WC star encounters the cooler OB stellar wind is referred to as the colliding wind region (or CWR) characterized by high gas densities and temperatures. The relation between dust formation and CWRs was first demonstrated by Williams et al. (1990) for the archetypal periodic dust maker WR~140 and dust formation monitoring in the infrared (IR) has strengthened this connection (Monnier at al. 2002, Williams et al. 2009). Indeed, the dust formation period of $\sim$ 7.9 years in WR~140 coincides with the periastron passages on the highly eccentric orbit of the binary system. At periastron, shocked X-ray and radio-bright gas has densities greatly enhanced in the CWR (Usov 1991) which will facilitate dust condensation (see Sections 4 and 6). Permanent dust makers are most likely part of binary systems as their periodic counterparts, but with rather small circular orbits. The prototype of such a subclass is the WC9 star WR~104 or so-called 'pinwheel system' for near-IR aperture-masking interferometry observations by Tuthill et al. (1999, 2008) reveal the long-range spiral structure of dust emission, as shown in Figure 1. Evidence for a time lag of $\sim 90$ days in the IR dust emission with respect to periastron indicates that dust condensation takes place downstream from the stagnation point of the CWR where temperatures are cool enough for dust to nucleate and condense (Williams 1999). 
 
\begin{figure}[!ht]
\includegraphics[scale=0.2, width=330pt]{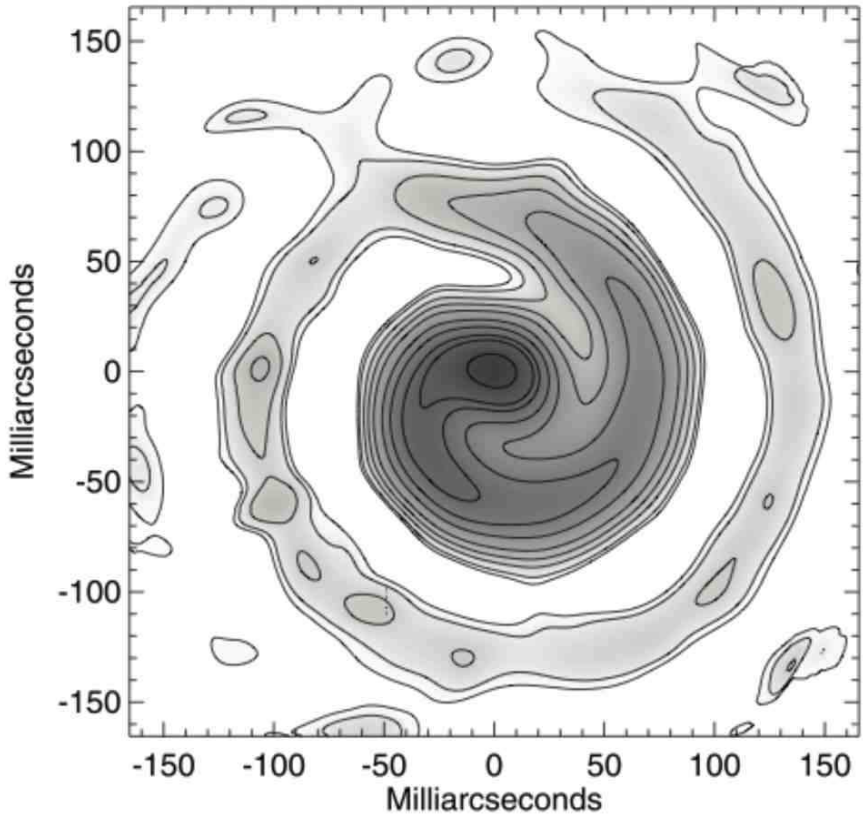}
\caption{Stacked composite image at 2.27 $\mu$m of the dust in the WR~104 pinwheel system from Tuthill et al. (2008).}
\end{figure}
\section{Evidence for dust in Supernovae close by and at high redshift}
 The explosion of SN1987A in the Large Magellanic Cloud provided for the first time evidence for molecule and dust formation in a SN. The 2.3 $\micron$ first overtone and 4.6 $\micron$ fundamental band of CO were first observed as early as 112 days post-explosion (Catchpole \& Glass 1987, Spyromilio et al. 1988) while excess emission between 8 $\micron$ to 9.5 $\micron$ was ascribed to vibrationally excited SiO from day 160 to day 517 after explosion (Aitken et al. 1988, Roche et al 1991). More recently, CO was detected in the Type II SNe SN1995ad (Spyromilio \& Leibundgut 1996), SN1998s (Gerardy et al. 2000) and SN2002dh (Pozzo et al. 2006), whilst SiO was detected by Kotak et al. (2006) in SN2005af. As to dust in SN1987A, a far-IR excess in the bolometric light curve, along with a sharp drop in the optical light curve, and an asymmetry in the optical emission lines proved to be due to solid forming in the ejecta about 400 days post-explosion (Lucy et al. 1989, Wooden et al. 1993).  According to Lucy et al., a mix of small silicate and amorphous carbon grains best reproduced IR observations. A lower limit to dust mass of $\sim$ 10$^{-4}$ \Ms~ was derived at 775 days post-explosion by Wooden et al. (1993). Recent modeling by Ercolano et al. (2007) for clumpy SN ejecta shows that the dust mass increases by a factor $\sim$ 10 when a clumpy ejecta is considered. They estimate that between 2 $\times 10^{-4}$ and 1.3 $\times 10^{-3}$ \Ms~of carbon grains condense in SN1987A at day 615 post-explosion. Dust forming in situ in more recent SN ejecta is too observed, specifically in SN2003gd (Sugerman et al. 2006) and SN2004et (Sahu et al. 2006). However, evidence for large quantities of dust condensing in local SNe is still sparse. 
 
Conversely, large amounts of dust are inferred to explain the reddening of background quasars and Lyman $\alpha$ systems at high redshift (Pettini et al. 1994). Observation at far-IR and submillimeter wavelength of J1148+5251, a hyper-luminous IR galaxy located at z = 6.4 implies the presence of $\sim 2 \times 10^8$ \Ms~at an age of $\sim$ 400 Myr (Bertoldi et al. 2003, Dwek et al. 2007). Because of their evolutionary time comparable to the age of the universe at z$>$ 6, low-mass stars cannot be the major dust providers as they are in our galaxy. The first stars ($\equiv$~Pop. III stars) are believed to be very massive owing to the lack of metals and consequent cooling and low-mass fractionation in primordial clouds. They evolve without losing mass until they explode as PISNe (Umeda \& Nomoto 2002, Heger \& Woosley 2002). Therefore the only possible place for dust to condense is in the expending ejecta resulting from the explosive end of Pop.~III stars.

 \section{Dust formation models for SNe and WC winds and the chemical kinetic approach} 
  
\begin{figure}[!ht]
\includegraphics[scale=0.25, width=330pt]{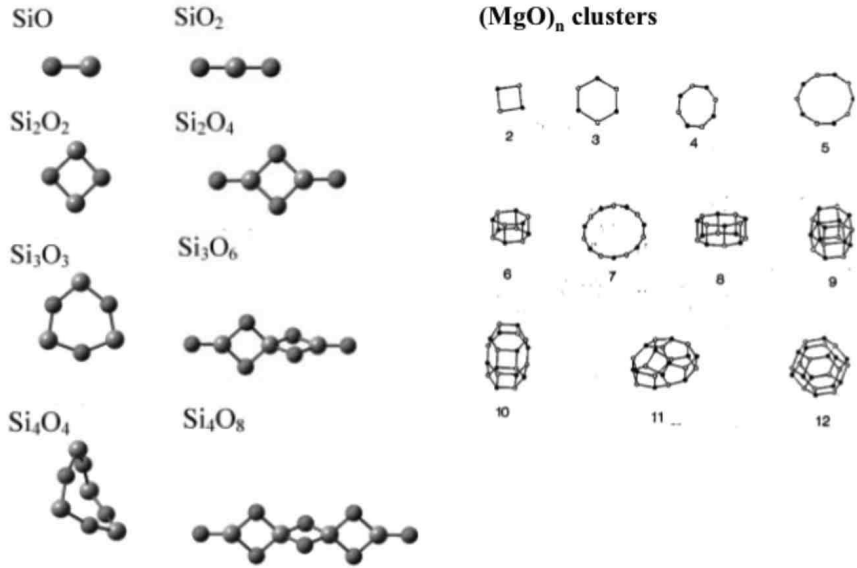}
\caption{Molecular precursors to forsterite dust condensation. Left: small silicon oxide clusters (SiO)$_n$ and (SiO$_2$)$_n$ for $n=1-4$ (Lu et al. 2003); Right: small magnesium oxide clusters up to $n=12$ (Ziemann \& Castleman 1991).}
\end{figure}

 The modeling of dust condensation in supernovae ejecta was first motivated by the dust detection in SN1987A. A first attempt was made by Kozasa et al. (1989) using a classical homogeneous nucleation theory. Based on thermal equilibrium analysis, they found that graphite and a mixture of forsterite, corundum and magnetite could form in C-rich and O-rich zones of the He-core, respectively. From the analysis of meteoritical inclusions bearing the isotopic signature of Type II Supernovae, Clayton et al. (1997) deduced that carbon-based dust (diamond and graphite) was the prevalent solid to form in the ejecta along with silicon carbide SiC,  and silicon nitride Si$_3$N$_4$. More recent meteoritical studies report the extraction of silicate grains as well  (Zinner 2006). Condensation of carbon solids in the H-free environment  of a local CCSN through formation of small carbon chains is modeled by Clayton et al. (1999) assuming chemistry at steady state in the O-rich ejecta, and carbon atoms produced by CO dissociation resulting from collisions with  $\gamma$-ray-induced Compton electrons.  At high redshift, Todini \& Ferrara (2001) model the dust formation in fully-mixed CCSN ejecta of various metallicities using too a classical nucleation theory and taking into account the formation of CO and SiO at steady state.  Schneider et al. (2004) extend their study to high-mass stellar progenitors. Finally, Nozawa et al. (2003) model solid formation in Pop. III progenitor SNe using Kozasa et al. (1989) formalism. All high redshift studies find that PISN ejecta  do form dust in significant quantities representing 15 to 30 \% of the progenitor mass. In view of the importance of identifying dust makers in the early universe, they conclude that PISNe can well account for the amount of dust conjectured at z $>$ 6.     

However, Donn \& Nuth (1985) already questioned the appropriateness of classical homogeneous condensation theory applied to dust condensation in circumstellar environments. Indeed, a concept such as 'surface tension' does not apply to very small clusters and equilibrium does not prevail in dynamical stellar winds or SN ejecta. A kinetic approach describing the formation of solids from the gas phase through nucleation and condensation was first adopted by Frenklach \& Feigelson (1989) and Cherchneff et al.~(1992) to model carbon dust formation in the winds of evolved AGB stars.  More recently, Cherchneff \& Lilly (2008, hereafter CL08) and Cherchneff \& Dwek (2009, hereafter CD09) describe the chemical kinetics of supernova ejecta at high redshift. They construct a chemical reaction network relevant to the formation of molecules and small clusters. The network is applied to various fully microscopically-mixed ejecta of massive progenitors as well as unmixed ejecta of PISNe and CCSNe. One clear conclusion is that steady state does not prevail in the ejecta of SNe in general for the post-explosion times characteristic of dust formation.  
 
 As for WC winds, dust condensation though observed has not yet been fully modeled. A first attempt to understand possible mechanisms whereby dust can grow in those massive, hot winds is proposed by Zubko (1998). Carbon dust growth is described by the addition of carbon ions to pre-existing positively-charged small carbon seeds and applied to spherically symmetric winds. Zubko finds that large carbon grains of 100 - 200~\AA~radius can form. A kinetic approach is proposed by Cherchneff \& Tielens~(1995) and Cherchneff et al.~(2000) who study the formation of small carbon chain precursors and their ions in the H-free, He-rich outflows of WC stars coupled to the formation of molecules like CO, O$_2$ or SiO which are relevant to the initial chemical composition of the wind. They find that the formation of small carbon chains is ineffective and requires higher gas densities than those of a spherically-symmetric steady wind.  As discussed in Section 6, the required density enhancement can be found in WC CWRs. 

The chemical processes whereby dust nucleates in massive circumstellar winds and SN ejecta are manifold and encompass high- and low-temperature chemistry for the gas spans large ranges of density and temperature. Here, we briefly discuss potential condensation routes for two major condensates in SN ejecta, carbon and forsterite (Mg$_2$SiO$_4$), the magnesium-rich end member of the silicate olivine. For carbon dust, there exist two routes dependent of temperature and the amount of hydrogen present in the gas. Firstly, a low-temperature (900-1700~K) channel in the presence of hydrogen involves the nucleation of polycyclic aromatic hydrocarbon (PAH) from pre-existing hydrocarbons such as acetylene (C$_2$H$_2$) and its isomers, propargyl (C$_3$H$_3$) and vinylacetylene (C$_4$H$_4$). The fundamental growing steps are (Frenklach \& Feigelson 1989, Cherchneff et al. 1992, Richter \& Howard 2002):

\begin{enumerate}
 \item Formation of a radical site through hydrogen abstraction and growth through acetylene addition,
 \item Closure of the first aromatic ring ($\equiv$ benzene C$_6$H$_6$) and subsequent growth and ring closure according to 1. 
\item  Once graphene sheets are formed, stacking of the sheets through long-range van der Waals forces and coagulation resulting in amorphous carbon grains. 
\end{enumerate}

The bottleneck in the above processes is the closure of the first ring which happens at temperatures in the range 900-1700~K.  Secondly,  a high-temperature (3000 - 4000~K) pathway via the formation of small and long branched carbon chains, monocyclic rings, rings with attached polyynes, bended aromatic structures resulting from the inclusion of pentagonal rings, and possible fullerene (C$_{60}$) closed cage structures (Curl \& Haddon 1993, Kroto et al. 1993, Cherchneff et al. 2000, Irle et al. 2006). When hydrogen is present, incomplete closure of fullerene cages occurs as observed in high temperature laser-induced pyrolysis of hydrocarbons (J{\" a}ger et al. 2009). 

For forsterite, the nucleation and condensation steps must include inhomogeneous processes. It is observed in the laboratory that forsterite forms from coagulation of silica (SiO$_2$) and periclase (MgO) clusters (Kamitsuji et al. 2005). Those precursors nucleate from the oxidation of magnesium vapor and evaporation of (SiO) clusters in dioxygen. Therefore, the nucleation of forsterite involves small molecular clusters for which structures have been theoretically derived by ab Initio calculations (Ziemann \& Castleman 1991, Lu et al. 2003). As an example, silica and periclase dust precursors are illustrated in Figure 2.  

\section{Chemical kinetic modeling of Pop~III Supernova ejecta}

\begin{figure}[!ht]
\includegraphics[scale=0.25, width=330pt]{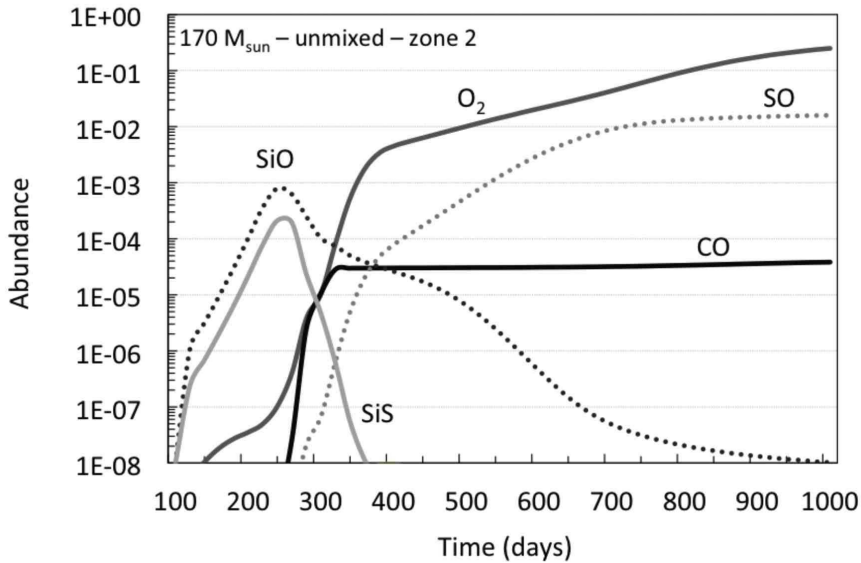}
\caption{Molecular abundances as a function of total gas number density for Zone 2 of the unmixed PISN ejecta with a 170 \Ms~progenitor (from Cherchneff \& Dwek 2009).  }
\end{figure}

\begin{table}[!ht]
  \begin{center}
  \caption{Molecular and dust masses ejected at 1000 days post-explosion for the unmixed 170 \Ms~and 20 \Ms~progenitors.}
 \begin{tabular}{lcc}
 \noalign{\smallskip}
 \noalign{\smallskip}
 \hline
 \noalign{\smallskip}
 Progenitor Mass& 170 M$_\odot$ &20 M$_\odot$ \\ 
  \noalign{\smallskip}
 \hline
  \noalign{\smallskip}
 {Molecules} & 37.1 M$_\odot$& 1.2 M$_\odot$\\ 
  \% of  Progenitor Mass  & 21.8 & 6.0 \\
  {Dust upper limit} & 5.5 M$_\odot$& 0.2 M$_\odot$\\ 
   \% of  Progenitor Mass & 3.2 & 1.0 \\
 \noalign{\smallskip}
  \hline
  \end{tabular}
\end{center}
\end{table}

As  previously mentioned, chemical ejecta of PISNe modeled with a chemical kinetic approach have been recently presented by CL08 assuming fully-microscopically mixed and unmixed ejecta. Mixing is likely to occur during SN explosion for the instability of the nickel bubble resulting in the development of Rayleigh-Taylor instabilities (M{\"u}ller et al. 1991, Kifonidis et al. 2003) but such a mixing is macroscopic rather than microscopic. Therefore, fully-microscopically mixed ejecta are highly unrealistic and their related modeling results only hint at the important processes at play in the ejecta. Of particular interest is the formation of a significant molecular phase in PISN ejecta which depletes part of the heavy elements available to dust nucleation from the gas. More recent and complete primordial SN models based on a similar approach are presented in CD09. They find that a 170 \Ms~unmixed PISN ejecta is able to synthesize a molecular phase of $\sim$ 37 \Ms. Prevalent chemical species are O$_2$, SiS, CO and SO for the unmixed model, equivalent to $\sim$ 70, 20, 8 and 2 \% of the molecular content, respectively. Figure 3 illustrates the molecular abundances in Zone 2 of the 170 \Ms~unmixed ejecta (i.e., the mass zone comprised between 20 and 40 \Ms~in the 82 \Ms~He-core). Table 1 summarizes the molecular budget found for a zero-metallicity 170 \Ms~PISN and for a 20 \Ms~CCSN. Upper limits for the dust ejecta content are derived by assuming that the total amount of dust precursors per type of grains will at the end coagulate and be transformed into a solid form. Dust precursors include small silica, iron sulfide, and periclase clusters, and carbon chains when CO thermal fragmentation along with CO conversion to C$_2$ is included into the model. As already stressed by CL08, carbon chains command carbon dust nucleation whilst the aromatic route stays always a minor formation channel, even in the presence of hydrogen. Temperature is the clincher, i.e., carbon chains form prior to aromatics and trigger nucleation via monocyclic rings in those hot environments. The dust upper limits listed in Table 1 are smaller than dust mass values derived by Nozawa et al. (2003) for similar unmixed SN models, pointing to the fact that the chemistry and the synthesis of molecules within the ejecta are a bottleneck to dust formation. 

\section{Dust formation in CWRs}
No complete dust formation model including the chemical description of nucleation and condensation of solids exists to date for WC stars. On the other hand, theoretical descriptions of colliding wind hydrodynamics have been performed by several groups. Stevens et al. (1992) show that a Kelvin-Helmholtz instability forms at the contact discontinuity between two adiabatic winds whereas a dense, thin-shell instability forms when one of the wind or both winds experience strong cooling. Walder \& Folini (2003) present 3D hydrodynamical simulations of CWRs and argue that the high densities requested for dust formation are found in the contact zone only. Pittard (2007) too considers the hydrodynamics of CWRs assuming the adiabatic winds are clumpy. He shows that initial structures/clumps in the winds are smoothed and that gas densities in the turbulent CWR are greatly enhanced, with mixing from both winds. We see from the above section that the carbon-dust formation channel in CWRs is definitely of high-temperature type, even if some hydrogen mixing occurs. It should thus proceed through formation of carbon chains, rings and intermediate fullerene caged species at gas temperatures as high as 4000 K and gas densities of $\sim 10^{10}$ cm$^{-3}$. Those conditions may be found in the CWR downstream from the central region. However, the nucleation of carbon clusters involves a molecular phase (possibly CO, CS, O$_2$ and small C$_n$ chains) bathed in strong radiation fields. How those chemical species withstand the inhospitable CRW conditions is still an open question. Observations of key molecules such as CO and CS coupled to new chemical/dynamical modeling of dust formation in those complex and fascinating environments are thus urgently needed.

\end{document}